 \documentclass{emulateapj}

\newcommand {\apgt} {\ {\raise-.5ex\hbox{$\buildrel>\over\sim$}}\ }
\newcommand {\aplt} {\ {\raise-.5ex\hbox{$\buildrel<\over\sim$}}\ }

\newcommand{\etal}{et al.~}%

\def\sun{\hbox{$\odot$}}
\def\cm3{cm$^{-3}$}

\def\rsun{R$_{\odot}$}

\def\mo{M$_{\odot}$}

\def\beq{\begin{equation}}
\def\eeq{\end{equation}}

\def\sles{\lower2pt\hbox{$\buildrel {\scriptstyle <}
   \over {\scriptstyle\sim}$}}
\def\sgreat{\lower2pt\hbox{$\buildrel {\scriptstyle >}
   \over {\scriptstyle\sim}$}}

\slugcomment{To Appear in ApJ Letters 2008 August 20}
\shorttitle{Probing Shock Breakout in Two SNLS Type II-P SNe with \textsl{GALEX}}
\shortauthors{Gezari et al.}

\begin{document}

\title{Probing Shock Breakout with Serendipitous \textsl{GALEX} Detections \\ of Two SNLS Type II-P Supernovae}

\author{Suvi Gezari,\altaffilmark{1}
Luc Dessart,\altaffilmark{2,3}
St\'ephane Basa,\altaffilmark{4}
D.~Chris Martin,\altaffilmark{5}
James D.~Neill,\altaffilmark{5}
S.~E. Woosley,\altaffilmark{6}
D.~John Hillier,\altaffilmark{7}
Gurvan Bazin,\altaffilmark{8}
Karl Forster,\altaffilmark{5}
Peter G.~Friedman,\altaffilmark{5}
J\'er\'emy Le Du, \altaffilmark{9}
Alain Mazure,\altaffilmark{4}
Patrick Morrissey,\altaffilmark{5}
Susan G.~Neff,\altaffilmark{10}
David Schiminovich,\altaffilmark{11}
and
Ted K.~Wyder\altaffilmark{5}
}
\altaffiltext{1}{Department of Physics and Astronomy,
        Johns Hopkins University,
        3400 North Charles Street,
        Baltimore, MD 21218 \email{suvi@pha.jhu.edu}}

\altaffiltext{2}{Steward Observatory, 933 North Cherry Avenue,
                 Tucson, AZ 85721}

\altaffiltext{3}{Department of Astrophysical Sciences, Princeton University,
                 Princeton, NJ 08544}

\altaffiltext{4}{Laboratoire d'Astrophysique de Marseille,
        13376 Marseille Cedex 12, France}

\altaffiltext{5}{California Institute of Technology,
        MC 405-47,
        1200 East California Boulevard,
        Pasadena, CA  91125}

\altaffiltext{6}{Department of Astronomy and Astrophysics,
                 University of California, Santa Cruz,
                 CA 95064}

\altaffiltext{7}{Department of Physics and Astronomy,
                 University of Pittsburgh, 3941 O'Hara Street,
                 Pittsburgh, PA 15260}

\altaffiltext{8}{DSM/DAPNIA, CEA/Saclay, 91191 Gif-sur-Yvette Cedex, France}

\altaffiltext{9}{CPPM, CNRS-IN2P3 and Universitat Aix-Marseille II, Case 907, 13288 Marseille Cedex 9, France}

\altaffiltext{10}{Laboratory for Astronomy and Solar Physics,
     NASA Goddard Space Flight Center,
       Greenbelt, MD  20771}

\altaffiltext{11}{Department of Astronomy,
     Columbia University,
         New York, NY  10027}

\begin{abstract}
  We report the serendipitous detection by \textsl{GALEX}
of fast ($<$1\,day) rising ($\apgt$1\,mag) UV emission from two
Type II plateau (II-P) supernovae (SNe) at $z$=0.185 and 0.324 discovered by the Supernova
Legacy Survey.
Optical photometry and VLT spectroscopy
2 weeks after the \textsl{GALEX} detections link the onset of UV emission
to the time of shock breakout.
Using radiation hydrodynamics and non-LTE radiative transfer simulations, and starting
from a standard red supergiant (RSG; Type II-P SN progenitor) star evolved self-consistently
from the main sequence to iron core collapse, we model the shock breakout phase and
the 55 hr that follow.
The small scale height of our RSG atmosphere model suggests that the breakout signature
is a thermal soft X-ray burst ($\lambda_{\rm peak}\sim$90\AA) with a duration of $\sles$ 2000\,s.
Longer durations are possible but require either an extended and tenuous non-standard envelope,
or an unusually dense RSG wind with $\dot{M} \sim 10^{-3}$\,\mo\,yr$^{-1}$.
The \textsl{GALEX} observations miss the peak of the luminous ($M_{\rm FUV}\approx-20$)
UV burst but unambiguously capture the rise of the emission and a subsequent 2 day long plateau. The postbreakout, UV-bright plateau is a prediction of
our model in which the shift of the peak of the spectral energy distribution (SED) from
$\sim$100 to $\sim$1000\AA\ and the ejecta expansion both counteract the decrease in
bolometric luminosity from $\sim$10$^{11}$ to $\sim$10$^9$\,L$_{\odot}$ over that period.
Based on the observed detection efficiency of our study we make predictions for
the breakout detection rate of the \textsl{GALEX} Time Domain Survey.
\end{abstract}


\section{Introduction \label{intro}}

In recent years, with the advent of sensitive $\gamma$-ray and X-ray detectors,
high-energy radiation from stellar objects has been
observed routinely. This includes in particular the long-soft $\gamma$-ray bursts (GRBs),
associated at present with a subset of hydrogen- and helium-deficient (Type Ic) SNe (Woosley \& Bloom 2006).
Their duration of $\sgreat$ 2\,s suggests
a compact progenitor with a $\sim$1\,s light-crossing time, a Wolf-Rayet (WR) star.
Observations have revealed that there is in fact a continuum from $\gamma$-ray to X-ray signatures
in association with such SNe. Hence, some events are related to GRBs in association
with relativistic ejecta, while others are associated with X-ray flashes and the SN shock breaking out into
the progenitor's dense wind, as in the Type Ic SN 2006aj and the Type Ib SN 2008D
(Campana et al. 2006; Waxman et al. 2007; Soderberg et al. 2008.; Li 2007, 2008).
In such SNe, the subsequent thermal emission from the cooling stellar envelope has also been
observed (e.g., SN Ic 1998bw: Galama et al. 1998; SN Ib/c 1999ex: Stritzinger et al. 2002; SNe 2006aj and SN 2008D).

In contrast to Type Ib/c SNe, hydrogen-rich (Type II) SNe originate from main-sequence objects
in the mass range 8--25\,\mo,
with iron cores at the time of collapse buried under an envelope of $\sim$6--20\,\mo.
The envelope light crossing time is $\sim$0.65 $R_3$\,hr ($R_3$ is the progenitor radius
in 10$^3$\,\rsun) and varies from about 5 minutes for a blue supergiant (BSG) progenitor to 1 hr
in the largest RSGs (Levesque et al. 2005).
In Type II SNe, these two properties prevent a central-engine from powering relativistic ejecta and
from producing their associated non-thermal $\gamma$-ray/X-ray radiation.
Hence, a burst of high-energy radiation from a Type II SN is most likely associated with shock emergence
at the surface of the progenitor and its immediate aftermath.

The shock-breakout phenomenon has been studied theoretically (Falk 1978; Klein \& Chevalier 1978),
with specific attention to the BSG progenitor of SN 1987A (Ensman \& Burrows 1992; Blinnikov et al. 2000)
or Type Ib/c SNe (Blinnikov et al. 2002). When the radiation-dominated SN shock wave travels outward through
the envelope and reaches regions of moderate optical depth (i.e., $\tau$ of a few tens),
a radiative precursor initiates the brightening of the object leading to a brief soft X-ray burst.
This precursor starts when the radiative diffusion time ($t_{\rm diff} \sim 3 \tau H_\rho / c$;
$H_{\rho}$ is the atmospheric scale height) becomes shorter than the shock-travel time to the surface,
located at $\tau\sim$1. The envelope structure and extent are two key ingredients controlling the properties of
the radiative precursor. In the absence of a wind, the expected durations
are $\sim$1\,s in a WR star (Blinnikov et al. 2002), $\sim$100\,s in a BSG (Ensman \& Burrows), and
$\sim$1000\,s in a RSG (Klein \& Chevalier 1978). A very optically-thick wind, with a scale height comparable to
the stellar radius, may, however, dramatically increase the diffusion time, and consequently the precursor duration.

In this Letter, we report the serendipitous detection with \textsl{GALEX} of {\it rising}
UV emission in two Type II-P SNe discovered by the Supernova Legacy Survey (SNLS),  SNLS-04D2dc
(also reported by Schawinski \etal 2008)
and SNLS-06D1jd, and associate both events with the breakout phase.
This contrasts with previous early UV observations of Type II SNe, i.e., the Type II-pec
SN 1987A (Hamuy et al. 1988), the Type IIb SN 1993J (Schmidt et al. 1993; Richmond et al. 1994), and
the Type II-P SN 2006bp (Immler et al. 2007; Dessart et al. 2008, hereafter D08), in which they were all {\it fading} in the UV, since the SN ejecta had already begun their long-term cooling phase through expansion and radiation.
In \S 2, we describe the \textsl{GALEX} and SNLS observations.
In \S \ref{sect_model}, we present our numerical simulations of
the breakout and the early post-breakout evolution in a standard RSG progenitor,
and compare these results to observations. In \S \ref{sect_discussion}, we conclude and
make predictions for the detection rate by the \textsl{GALEX} Time Domain Survey (TDS).
Throughout this Letter we adopt $H_{0}= 70$ km s$^{-1}$ Mpc$^{-1}$, $\Omega_{\rm M}$=0.3,
and $\Omega_{\Lambda}$=0.7.

\section{Observations} \label{sect_obs}

We conducted a search for early UV emission in Type II SNe
by matching SNLS SN candidates with UV sources in overlapping \textsl{GALEX} Deep Imaging Survey (DIS) (Martin \etal 2005) fields.  We selected SN candidates that had $\sim 1.5$ ks \textsl{GALEX} DIS observations within 30 days before the optical SN candidate discovery, yielding a sample of 293 SN candidates with 1 to 77 \textsl{GALEX} observations within that time period.
Two variable UV sources
were detected with a 3$\arcsec$ matching radius to the position of SN candidates
SNLS-04D2dc (R.A. = 10h 00m 16.681s, decl. = $+02\degr 12\arcmin 18.52\arcsec$ [J2000.0])
and SNLS-06D1jd (R.A. = 02h 27m 36.189s, decl. = $-04\degr 31\arcmin 56.62 \arcsec$ [J2000.0]).
The \textsl{GALEX} photometry was measured using aperture magnitudes
with $r_{ap}=6\arcsec$ and a centroid fixed to that of the highest signal-to-noise (S/N)
detection, with an aperture correction to
account for the fraction of energy enclosed in the aperture of $m_{tot}-m_{ap} = -0.15$ and $-0.23$ mag
in the FUV  ($\lambda_{\rm eff}$=1539\AA) and  NUV ($\lambda_{\rm eff}$=2316\AA), respectively (Morrissey \etal 2007).

SNe SNLS-04D2dc and SNLS-06D1jd were discovered on 2004 March 16 and 2006 December 12
by the SNLS real-time difference imaging pipeline (Astier \etal 2006; Sullivan \etal 2006).
Follow-up VLT-FORS1 spectroscopy was obtained on 2004 March 19 (JD 2,453,083.5) and
2006 December 22 (JD 2,454,091.5),
corresponding to 21.2 and 23.4 days after the \textsl{GALEX} detection, respectively.
The SNe were classified as Type II-P SNe based on the identification of broad
H{\sc i} emission lines (most noticeably H$\beta$ and H$\gamma$ in SNLS-06D1jd, but also
H$\alpha$ in SNLS-04D2dc), and a $\sim$40\,day plateau in optical brightness.
The host galaxy redshifts are measured
from nebular emission lines and the spectra and photometry are fitted with templates
of a late-type galaxy at $z=0.185$ with $m_{i'}$=19.6 for SNLS-04D2dc and a starburst
galaxy at $z=0.324$ with $m_{i'}$=22.1 for SNLS-06D1jd.
Both VLT spectra match well the observations of SN 1999em at $\sim$17 days after breakout
(Dessart \& Hillier 2006), suggesting that the observed UV brightening
is contemporaneous to the breakout phase.

The host galaxies are detected in the FUV and
NUV with $m_{\rm FUV}$=23.42$\pm$0.07 (23.80$\pm$0.08)
and $m_{\rm NUV}$= 22.94$\pm$0.06 (23.51$\pm$0.07) for SNLS-04D2dc (SNLS-06D1jd) measured from deep co-added images
constructed from observations taken before the SN detection with $t_{\rm exp}$= 37.6\,ks (64.0\,ks).
We show the light curves of SNe SNLS-04D2dc in Figure 1 and SNLS-06D1jd in Figure 2,
including \textsl{GALEX} FUV and NUV photometry (after subtraction of the host
galaxy contribution) and
SNLS difference imaging photometry in the $g'$
($\lambda_{\rm eff}$=4872~\AA), $r'$ ($\lambda_{\rm eff}$=6282~\AA), $i'$ ($\lambda_{\rm eff}$=7776~\AA), and
$z'$ ($\lambda_{\rm eff}$=8870~\AA) bands.
Note the $\sgreat$12\,day gap between the UV and optical observations of both SNe.
In order to relate the UV and optical photometry, we create synthetic \textsl{GALEX}-
and SNLS-filter light curves using the multi-epoch non-LTE synthetic spectra
computed for the Type II-P SN 2006bp ($d = 17.5$\,Mpc; D08) and adjust to the
distance, redshift, and extinction of SNLS-04D2dc and SNLS-06D1jd.
For each, we adopt the onset of UV emission as the breakout time,
yielding JD 2,453,062.25$^{+0.10}_{-0.25}$ (JD 2,454,068.05$^{+0.01}_{-0.74}$)
for SNLS-04D2dc (SNLS-06D1jd).
For SN 2006bp, we adopt JD 2,453,834.2 (Quimby et al. 2007).
The optical colors during the plateau of both SNLS SNe are well fitted with a reddening of $E(B-V)\sim$0.1.
However, an offset of $+0.3$ ($-0.4$) for
SNLS-04D2dc (SNLS-06D1jd) was added to obtain the best match to the observations, suggestive of a difference
in intrinsic brightness of these SNe compared to SN 2006bp or an uncertainty in the extinction.

\section{Models and results}\label{sect_model}

  We model the UV and optical light curves from 0.05
to 55.6\,hr after shock breakout in three consecutive steps.
First, using the one-temperature Lagrangian radiation hydrodynamics code KEPLER
(Weaver et al. 1978),
the evolution at solar metallicity of a 15\,\mo\ main sequence star is computed at high resolution
($\Delta M \sim 10^{-6}$\,\mo\ at the surface) until iron core collapse.
At that time, the star is a RSG with a radius $R_{\star}$=865\,\rsun\ and a total mass of
12.58\,\mo\ (Woosley \& Heger 2007).
The star above a mass cut of 1.45\,\mo\ is then artificially exploded with a piston
to yield an asymptotic SN ejecta kinetic energy of 1.2$\times$10$^{51}$\,ergs typical of a core-collapse SN.\footnote{However, originally,
about half of the energy deposited by the shock wave is in the form
of heat and trapped-radiation (internal  energy), so that at shock breakout,
i.e., 1.47\,days after the piston trigger, the bulk of the SN ejecta in our model has not
yet reached its coasting velocity.}
We finally post-process a sequence of 20 KEPLER models with the non-LTE
non-relativistic steady-state code CMFGEN (Hillier \& Miller 1998; Dessart \& Hillier 2005a).
The solution for level populations and the radiation field at $\sim$10$^5$ frequencies
is, thus, iterated to convergence on the stationary hydrodynamical structure,
but allowing the temperature to relax in regions exterior to a Rosseland-mean optical depth
of $\sim$20. This ensures that our assumption of steady-state is consistent, i.e., the diffusion time at the
RSG surface being much less than the SN age. This is the first time the phase contemporaneous to
shock breakout is modeled with a non-LTE multi-angle multi-frequency radiative transfer code,
which allows for both line and continuum sources of opacities and treats
multiple species and ionization states, including H{\sc i}, He{\sc i}--{\sc ii},
C{\sc ii}--{\sc v}, N{\sc ii}--{\sc v}, O{\sc ii}--O{\sc vi}, Ne{\sc iv}--{\sc viii}, and Fe{\sc iv}--{\sc xi} and
corresponding to a total of $\sim$1000 levels.

The one-temperature treatment adopted in KEPLER forces the gas and the radiation to the same temperature,
and, thus, prevents our modeling of the radiative precursor. However, the small atmospheric scale height
$H_\rho \sim$0.01\,$R_{\star}$ in our RSG progenitor surface suggests that the precursor
will last $\sim$1000\,s, light-travel-time effects smearing the observed signal over $\sim$$R_{\star}/c\sim$2000\,s.
Longer durations are possible, but would require a larger atmospheric scale
height, perhaps reaching $H_\rho \sim R_{\star}$.
This may be produced by an extended low-density envelope, as adopted by \cite{sch08}, although
this appears incompatible with the RSG envelope structure computed with KEPLER, the 3D hydrodynamics simulations
of RSG envelopes (Freytag et al. 2002), and the quantitative spectroscopic analyses of RSG stars (Josselin \& Plez 2007).
Alternatively, this may be produced by an RSG in a super-wind phase, i.e., an RSG with an unusually high mass-loss rate
$\dot{M} \sim 10^{-3}$\,\mo\,yr$^{-1}$ compared to the more standard values of a few
$\times$10$^{-5}$\,\mo\,yr$^{-1}$ \citep{salasnich:99}.

In Figure \ref{fig_phot_prop}{\it a}, we show in black the evolution of the temperature $T_{\rm phot}$
and radius $R_{\rm phot}$, defined by $\int_{R_{\rm phot}}^\infty \kappa_{\rm t}dR$= 2/3,
where $\kappa_{\rm t}=\kappa_{\rm a}+\sigma$ and $\kappa_{\rm a}$ ($\sigma$) is the
absorptive (scattering) opacity.
$T_{\rm phot}$ reaches up to $\sim$1.35$\times$10$^5$\,K only $\sim$0.5\,hr after breakout, associated with
a brightening phase at constant radius (light-travel time effects and the radiative precursor
would yield a $\sim$2000\,s signal instead).
Note that at this time, the ``RSG'' has the surface temperature of a white dwarf
and radiates at 10$^{11}$ times the rate of the Sun!
The temperature, decreasing almost as quickly as it rose and reaching down
to 40,000\,K after $\sim$5\,hr, dominates the bolometric variation since mass shells
have hardly moved during this early phase.
At 55\,hr after breakout, $T_{\rm phot}\sim$20,000\,K,
and $R_{\rm phot}$ has increased by only a factor of 3.8 to $\sim$2.3$\times$10$^{14}$\,cm.
KEPLER predicts a steep density distribution at the photosphere, scaling as 1/$R^{200}$ at breakout
to 1/$R^{50}$ 55\,hr later, a property also inferred from line profile morphology in SN 2006bp by D08.
With electron scattering dominating the opacity, the emergent SN flux suffers dilution
(Dessart \& Hillier 2005b), reflecting better the temperature at the thermalization radius $R_{\rm th}$
(defined by $\int_{R_{\rm th}}^\infty \sqrt{\kappa_{\rm a} \kappa_{\rm t}}dR$= 2/3) rather
than at the photosphere (Fig.~3{\it a}).
These temperatures differ by up to a factor of 2--3 at all times, and the SED is much harder than the blackbody
distribution at $T_{\rm phot}$, with a peak (blue curve in Fig.~\ref{fig_phot_prop}{\it a})
at $\sim$90~\AA\, (900~\AA) at 0.5\,hr (55.6\,hr) after breakout. The non-LTE synthetic spectra shown
in Figure \ref{fig_phot_prop}{\it b} illustrate these properties and reveal for the first time
the line features predicted under these breakout conditions (these characteristics will be discussed
in a future paper).

\begin{figure*}
\plotone{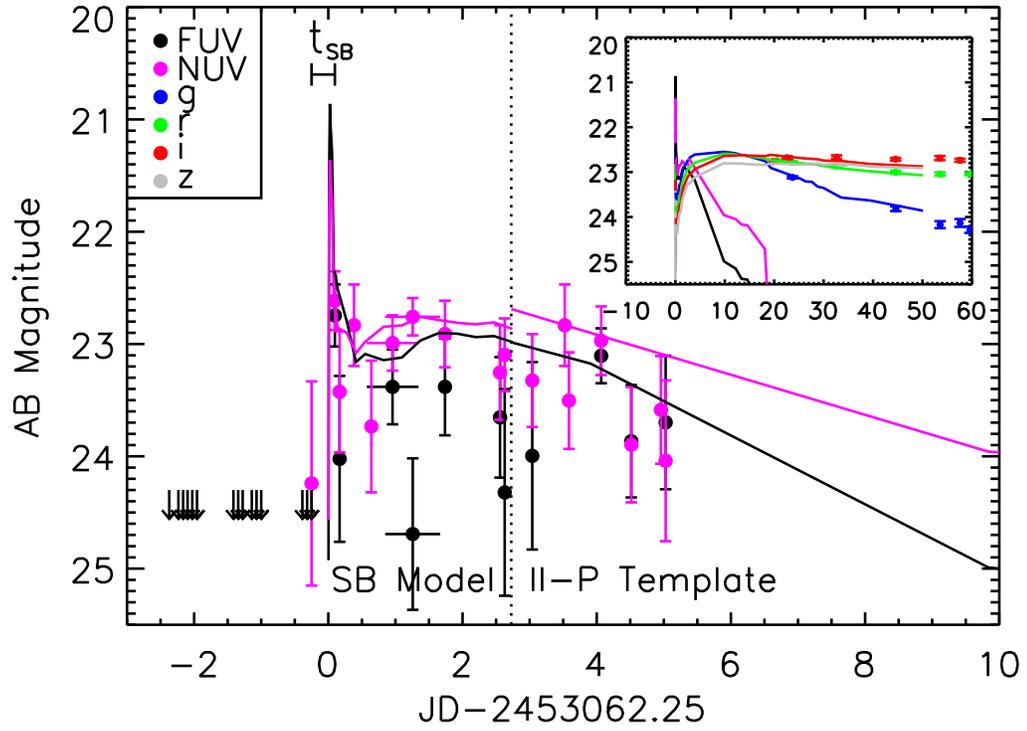}
\caption{
\textsl{GALEX} and SNLS ({\it inset}) light curves ({\it colored symbols}) of SNLS-04D2dc
($z=$0.185) shown relative
to the estimated breakout time ({\it horizontal bar}, labeled $t_{\rm SB}$)
and overplotted with synthetic magnitudes ({\it colored lines}; II-P Template:
scaled synthetic light curves of SN 2006bp [D08]; SB Model: shock-breakout model).
Horizontal error bars show the time coverage of photometry measured from coadded observations.
Solid arrows show 95\% upper limits.  The \textsl{GALEX} FUV detector was temporarily not operational
during the NUV observations around $t\sim3.5$ days.
\label{fig_GALEX_LCa}
}
\end{figure*}

\begin{figure*}
\plotone{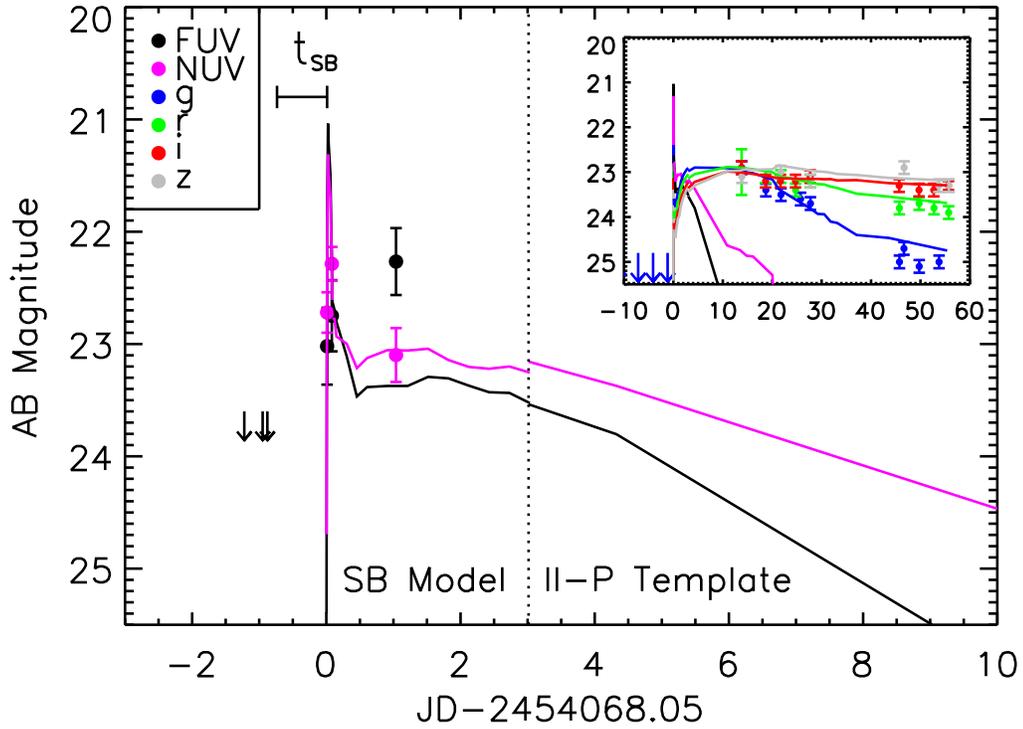}
\caption{
\textsl{GALEX} and SNLS ({\it inset}) light curves ({\it colored symbols}) of SNLS-06D1jd ($z=$0.324), shown relative
to the estimated breakout time ({\it horizontal bar}, labeled $t_{\rm SB}$),
and overplotted with synthetic magnitudes ({\it colored lines}; II-P Template:
scaled synthetic light curves of SN 2006bp [D08]; SB Model: shock-breakout model).
Solid arrows show 95\% upper limits.
\label{fig_GALEX_LCb}
}
\end{figure*}

In Figures \ref{fig_GALEX_LCa} and \ref{fig_GALEX_LCb}, we show the model \textsl{GALEX} and SNLS light curves created from
our synthetic spectra for SNLS-04D2dc and SNLS-06D1jd.
The CMFGEN models ($1.67$ hr $\le t \le 55.6$ hr)
are offset by  $-1.5$ ($-1.4$) mag for SNLS-04D2dc (SNLS-06D1jd) to match the SN 2006bp template
light curves ($t>$2 days) that were scaled to fit the optical observations, suggesting that our
models underestimate the observed luminosity by a factor of $\sim$4.  We find no significant variability in the photon data between the first and second halves of the first two $\sim 20$ minute observations during the rise of the UV emission from SNLS-04D2dc and SNLS-06D1jd, with $\Delta$m$< 0.5$ mag.  Although the lack of significant intra-observation variability and the $\sim$ 2 mag amplitude of
variability predicted by our models for the $\sles$1\,hr long breakout signature implies that the \textsl{GALEX} observations do not
coincide with the peak of the UV burst, the onset of the fast brightening UV emission
is suggestive evidence that we have detected the breakout phase
and constrains the time of shock breakout to an accuracy of 0.35 (0.75) days
for SNLS-04D2dc (SNLS-06D1jd), i.e. 7.1 hr (13.7 hr) in the SN rest frame.
Within the errors, the subsequent 2\,day long UV plateau is well reproduced by
the model,
although we note the poor match to the noisy FUV data in the plateau of SNLS-04D2dc and
to the FUV/NUV color of SNLS-06D1jd at $\sim$1\,day.
Future, better sampled, higher S/N UV light curves of breakout events in Type II-P SNe
will provide more accurate constraints, foster a better understanding, and, thus,
permit more definite conclusions on this ephemeral breakout phase and the
RSG envelope structure.

\section{Discussion and conclusion}
\label{sect_discussion}

\begin{figure}
\plottwo{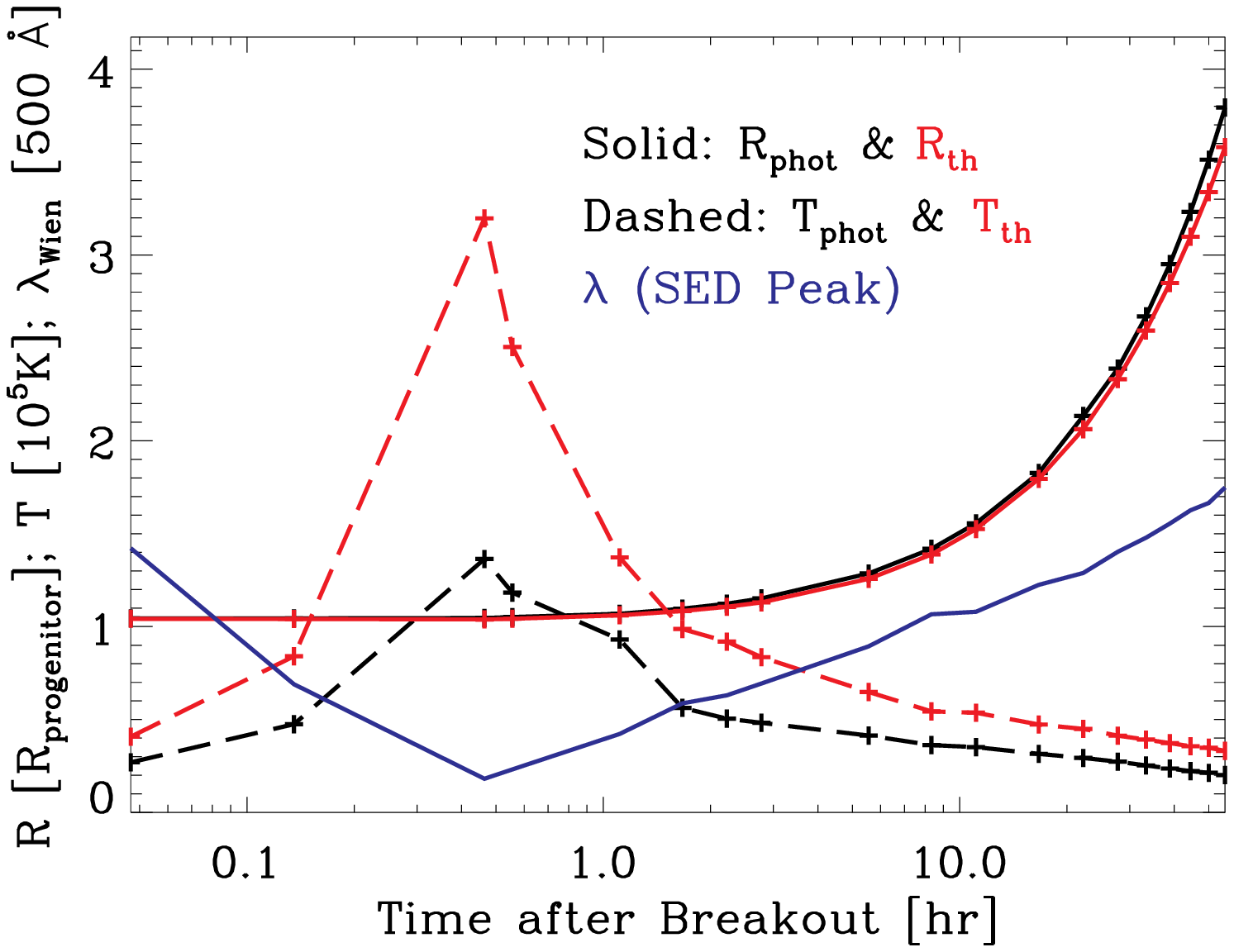}{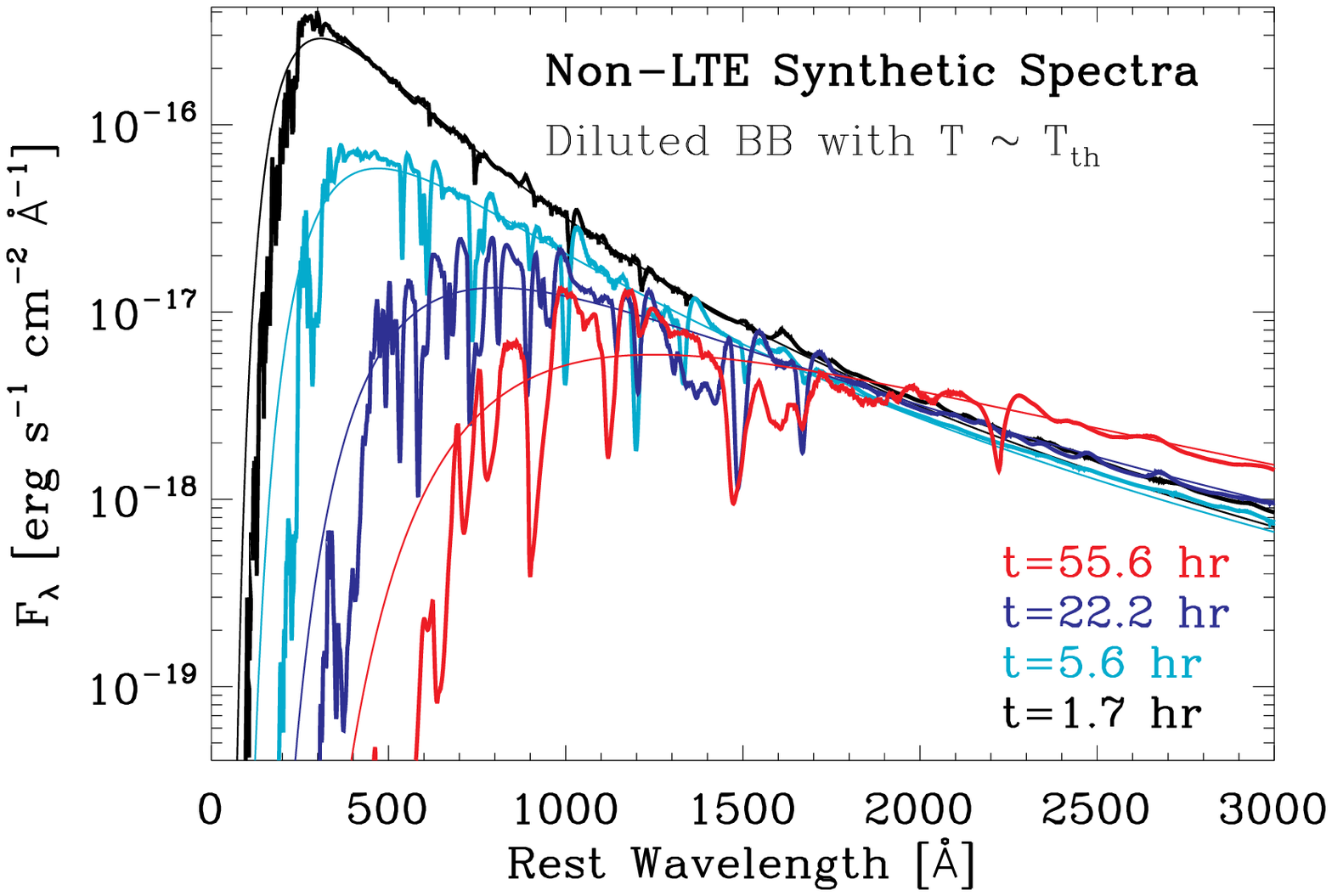}
\caption{{\it Left}: KEPLER/CMFGEN predictions for the post-breakout evolution of the wavelength of the SED peak
({\it blue line}; in units of 500\,\AA), together with the gas temperature (T$_{\rm phot}$; {\it dashed line})
and radius (R$_{\rm phot}$; {\it solid line}) at the photosphere ({\it black line}) and at the thermalization depth ({\it red line}).
{\it Right}: Multiepoch postbreakout synthetic non-LTE ({\it thick line}) and diluted blackbody ({\it thin line}; $T_{\rm BB} \sim T_{\rm th}$)
spectra, scaled to the luminosity distance of SNLS-06D1jd.
\label{fig_phot_prop}
}
\end{figure}

\textsl{GALEX} DIS serendipitously detected fast rising UV emission in two SNLS Type II-P SNe,
which we associate with the breakout phase in a RSG star.
The observed amplitude of the UV signal is smaller than that of the
$\sles$\,1\,hr long breakout burst predicted by our models, which
suggests that the observations did not capture the peak of the burst (we see
no strong variability within the 20 minute observations), although
our theoretical peak flux is likely overestimated by the one-temperature treatment in KEPLER.
The important conclusion from these models is that the observed UV brightening by $\sim$1\,mag
is predicted to occur only during this early, breakout, phase and thus is a direct probe of the
time of shock breakout in these SNe.
Moreover, we both observe and predict that the immediate post-breakout in the UV is followed by a plateau
phase of about 2 days, supporting a RSG-like progenitor (more compact progenitors,
e.g., BSGs, cool much faster after breakout, as seen for SN 1987A), and providing an interesting clock
to gauge the time since breakout.
In SN 2006bp, the SN was fading in the UV right from the start of data
collection, and a time since explosion of 1.7\,days was indeed inferred (D08).
The post-breakout UV plateau is the result of two competing effects, with the strong decrease in
bolometric luminosity compensated by the shift of the SED to longer wavelength as the ejecta
expand and cool (Fig.~\ref{fig_phot_prop}{\it a}). At maximum light, the peak of the SED in our models
is at 90~\AA\ (in the rest frame), and so the UV rise reveals only the evolution of the tail of the
SED, not that of the peak.

The volumetric rate of Type II-P SNe is close to that of Type Ia SNe \citep{dahl04}; thus, we use the measured fraction of Type Ia SNe in the total sample of SNLS SN candidates from Sullivan \etal (2006), $f_{Ia}=0.16$, to estimate the fraction of Type II-P SNe in our sample.  If we assume that the shock breakout occurs $15 \pm 3$ days before the optical discovery of the SN, then the number of candidates in our study that had a probability of at least one \textsl{GALEX} observation during the 2 day UV plateau following shock breakout is $n_{obs}=167$.  By contrast, the number of candidates that had a probability of having a \textsl{GALEX} observation that coincided with a 1\,hr long flash at the time of shock breakout is $\aplt$1.  We now estimate the number of Type II-P SNe observed during the UV-plateau phase as $n_{II-P} \approx f_{Ia}n_{obs} \sim 27$, and get a detection efficiency of $f_{det} = 2/n_{II-P} \sim 0.07$.
\textsl{GALEX} TDS will operate in parallel with the
Pan-STARRS Medium Deep Survey (PS MDS), an optical time domain survey that will monitor 50 deg$^{2}$ night$^{-1}$ with a depth comparable to SNLS. \textsl{GALEX} TDS plans to monitor $\sim$7\,deg$^{2}$ of the PS MDS with cadences ranging from 98.6 minutes
(two consecutive orbits) to several days for $\sim$8 months of the year to a depth
of $m_{lim}\sim$23\,mag. PS MDS is expected to have a SN Ia detection rate of $r_{Ia} \approx 23$ month$^{-1}$ per 7\,deg$^{2}$ field (J.~Tonry 2008, private communication), which for our observed detection efficiency ($f_{det} \sim 0.07$) and $n_{months}=8$ implies a favorable detection rate of UV emission from the shock breakout phase of Type II-P SNe by \textsl{GALEX} TDS of $r_{det} \approx r_{Ia}n_{months}f_{det} \approx$ 10 yr$^{-1}$. 

\acknowledgements

We thank the anonymous referee for their helpful comments.
L.D. acknowledges support for this work from the SciDAC program of the DOE, 
under grants DE-FC02-01ER41184 and DE-FC02-06ER41452,
and from the NSF under grant AST-0504947. S.W. acknowledges support from
the SciDAC under grant DE-FC02-06ER41438 and also by NASA under grant NNG05GG28G.  S.G. and S.B. thank Bruno Milliard for his support during this study.  We gratefully acknowledge NASA's support for construction,
operation, and science analysis for the \textsl{GALEX} mission,
developed in cooperation with CNES of
France and the Korean MOST.
Based on observations obtained with MegaPrime/MegaCam, a joint
project of CFHT and CEA/DAPNIA, at the
CFHT which is operated by the NRC of
Canada, the INSU of the
CNRS of France, and the
University of Hawaii. This work is based in part on data products
produced at TERAPIX and the CADC as part of
the CFHT Legacy Survey, a collaborative
project of NRC and CNRS.  
Based on observations made with the ESO telescopes at the La Silla or Paranal Observatories
under proposal IDs 171.A-0486 and 176.A-0589.

\clearpage

\end{document}